\documentclass[prl,aps,twocolumn,groupedaddress,showpacs]{revtex4}
\usepackage{amssymb,amsmath,amsfonts,epsfig,graphicx,latexsym,bm,color}

\newcommand{\beq}{\begin{eqnarray}}
\newcommand{\eeq}{\end{eqnarray}}
\newcommand{\nn}{\nonumber}

\begin{document}
\title{Unconventional superfluid order in the $F$-band of a bipartite optical square lattice}
\author{Matthias \"{O}lschl\"{a}ger, Georg Wirth, and Andreas Hemmerich}
\affiliation{Institut f\"{u}r Laser-Physik, Universit\"{a}t Hamburg, Luruper Chaussee 149, 22761 Hamburg, Germany}
\date{\today}
\date{\today}

\begin{abstract}
We report on the first observation of bosons condensed into the energy minima of an $F$-band of a bipartite square optical lattice. Momentum spectra indicate that a truly complex-valued staggered angular momentum superfluid order is established. The corresponding wave function is composed of alternating local $F_{2x^3-3x} + i\,F_{2y^3-3y}$-orbits and local $S$-orbits residing in the deep and shallow wells of the lattice, which are arranged as the black and white areas of a checkerboard. A pattern of staggered vortical currents arises, which breaks time reversal symmetry and the translational symmetry of the lattice potential. We have measured the populations of higher order Bragg peaks in the momentum spectra for varying relative depths of the shallow and deep lattice wells and find remarkable agreement with band calculations.
\end{abstract}


\maketitle
\date{\today}
The physics of condensed matter systems is often determined by electrons, which possess orbital degrees of freedom involving an intrinsic anisotropy due to multiple orbital orientations. Orbital physics plays a central role for magnetism, superconductivity, and transport properties of rare earth and transition metal compounds including high $T_c$ cuprate systems or heavy fermion systems, which are subject of intense research since more than two decades \cite{Sig:91, Mae:04, Bus:07}. The advent of optical lattices (i.e. quantum gases arranged in synthetic lattices formed by light) has raised hopes that certain aspects of such systems could be studied in a precisely controlled environment without many of the complexities usually associated with material systems \cite{Lew:07,Blo:08}. Unfortunately, the wave function of bosons in their ground state according to a consideration by Feynman, referred to as the "no node theorem", is positive definite under very general circumstances \cite{Fey:72, Wu:09}. This appears to significantly limit the use of bosons for simulating many-body systems of interest. Fermions, on the other hand, are significantly harder to prepare in optical lattices \cite{Koe:05, Chi:06} and the realization of large filling factors required to access relevant orbital physics appears difficult, if not impossible. It has long been speculated, that the use of higher bands could be a way to explore the interplay between superfluidity and orbital physics also with bosons \cite{Isa:05,Liu:06,Kuk:06, Xu:07,Sto:08, Mar:09}. The development of clever new techniques has made it possible to populate higher bands \cite{Bro:05, Mue:07, And:07, Wir:10}. In Ref. \cite{Mue:07} atoms could be excited into the $P$-band of a two-dimensional (2D) optical lattice and short-lived ($< 1$ ms) coherence was established along specific directions. Cross-dimensional coherence could be established more recently in a bipartite lattice due to cross-dimensional tunneling junctions realized by local $S$-orbits in every second well \cite{Wir:10}. However, the $D$-and $F$-bands in optical lattices have remained practically unexplored by experiments.

In this article we report the first observation of bosons condensed in the $F$-band of a quasi 2D bipartite optical lattice. Full cross-dimensional coherence with a life-time on the order of 10 ms is established. The observed momentum spectra exhibit a characteristic pattern of sharp maxima, which are well explained by a complex-valued superfluid order parameter composed of alternating local $F_{2x^3-3x} + i\,F_{2y^3-3y}$-orbits and local $S$-orbits. A pattern of staggered local angular momenta and staggered vortical currents arises commensurable with the plaquettes of the lattice with the consequence of broken time-reversal symmetry. The proposed nature of the superfluid order is confirmed by evaluating the Bloch functions corresponding to the observed condensation quasi-momenta, which are derived via two-dimensional band calculations. We measured the population ratios between higher order and lowest order Bragg peaks in the momentum spectrum for varying relative depths of the shallow and deep lattice wells and find remarkable agreement with calculations based upon these Bloch functions. 

By crossing two optical standing waves derived from laser beams with 100 $\mu$m $1/e^2$-radius and a wavelength $\lambda = 1064\,$nm we produce a (quasi 2D) light shift potential  
\beq
\label{M.1}
 V(x,y) \equiv- \frac{V_0}{4} \, e^{-\frac{2z^2}{w_0^2}} |\, \eta \, \left( \,e^{i k x}  + \epsilon \,e^{-i k x}\right)  \qquad \qquad   \\ \nn 
 + \,\, e^{i \theta} \, \left(e^{i k y} + \epsilon \, e^{-i k y} \right)|^2  \, ,
\eeq
providing two classes of (tube-shaped) lattice sites (denoted as $\mathcal{A}$ and $\mathcal{B}$) arranged as shown in Fig.1(a). Here, $k \equiv 2\pi/\lambda$, $\eta$ is experimentally adjustable (around unity) and $\epsilon$ is fixed to $\approx 0.9$ due to imperfect reflection optics used in the experiment. For details we refer to Ref.\cite{Wir:10}. Adjustment of the parameter $\theta$ lets us tune the difference of the well depths between $\mathcal{A}$ and $\mathcal{B}$-sites. For $\theta < \pi/2$ the $\mathcal{A}$-sites are more shallow than the $\mathcal{B}$-sites and vice versa. A change of $\theta$ by $\pi/2$ can be obtained in less than 0.2 ms. 

\begin{figure}
\includegraphics[scale=0.3, angle=0, origin=c]{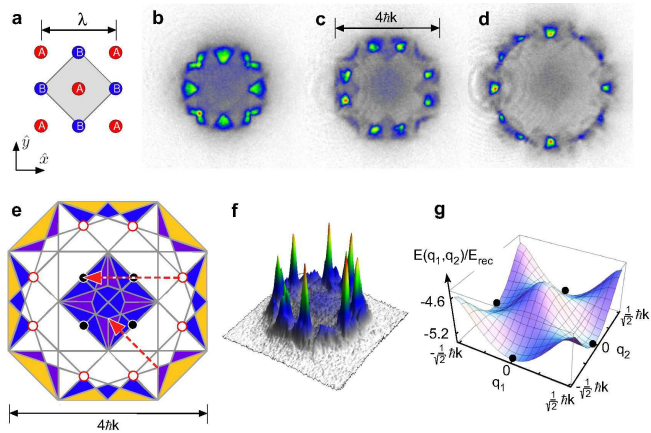}
\caption{\label{Fig.1} (a) The bipartite lattice comprises two classes of lattice sites denoted by $\mathcal{A}$ and $\mathcal{B}$. The grey area shows the Wigner Seitz unit cell of the $\mathcal{A}$-sublattice. (b), (c) and (d) Observed populations of Brillouin zones (BZs) after applying the population swapping procedure with final values of $(\theta / \pi, V_0/E_{\textrm{rec}}) = (0.61, 5.8 ), (0.66, 8.3 ), (0.69, 13.3)$, respectively. (e) Schematic of the 6th (blue and purple areas) and 7th (orange areas) BZs. The eight red circles mark the points, where the atoms are observed to gather in (c). In the center of the plot the first BZ is reconstructed by translations of subsets of the 6th BZ via reciprocal lattice vectors indicated by the red dashed arrows. (f) 3D representation of the image in (c). (g) energy surface of the 7th band with degenerate minima indicated by black disks.}
\end{figure}

Efficient population of excited bands is obtained by optimizing a population swapping procedure described in more detail in Ref.\cite{Wir:10}. Initially a Bose-Einstein condensate (BEC) of Rubidium ($^{87}$Rb) atoms is prepared and the lattice potential is ramped up within 80 ms to $V_0/E_{\textrm{rec}} = 16.6$ for a value $\theta < \pi/2$ such that the well depth of the $\mathcal{B}$-sites significantly exceeds that of the $\mathcal{A}$-sites ($E_{\textrm{rec}} \equiv \hbar^2 k^2/2m$ denotes the recoil energy with the atomic mass $m$). A ground state lattice is thus formed with most atoms residing in the deeper $\mathcal{B}$-wells. The large well depth yields nearly complete suppression of tunneling. Subsequently $\theta$ is rapidly changed (within 0.2 ms, which is shorter than the nearest neighbor tunneling time) to a final value $\theta_f$ above $\pi/2$ such that now the $\mathcal{A}$-wells are significantly deeper than the $\mathcal{B}$-wells. Finally, $V_0$ is adiabatically decreased during 0.6 ms to admit tunneling again. Optionally a 2 ms long phase is appended, where $\theta$ is adiabatically tuned to some desired value followed by a variable hold time.

In Fig.1(b),(c), and (d) the population of different bands is illustrated. To obtain these pictures, population swapping is carried out with $\theta_f$ and $V_0$ optimized for maximal population of the desired band. The lattice beam intensity is then exponentially decreased with a time constant of $430\,\mu$s and after 30 ms absorption imaging is applied. This yields images of momentum space, where the population $P[n,q]$ in the n-th band for some quasi-momentum $q$ and energy $E[n,q]$ is mapped to some point within the n-th Brillouin zone (BZ) related to $q$ by a reciprocal lattice vector. This mapping requires that $E[n,q]$ does not cross some other $E[n',q]$ during the adiabatic switch off process. As a consequence of such band crossings a part of the band population may be mapped to an adjacent BZ. The image in Fig.1(b) (with $\theta/\pi = 0.61$ and $V_0/E_{\textrm{rec}} = 5.8$) matches well with the 4th BZ, while in Fig.1(d) (with $\theta/\pi = 0.69$ and $V_0/E_{\textrm{rec}} = 13.3$) the 9th BZ is found to be the most populated. This directly indicates the numbers of the bands that have been populated to be $n=4$ and $n=9$, respectively. For the case of interest in this article in Fig.1(c), where $\theta/\pi = 0.66$ and $V_0/E_{\textrm{rec}} = 8.3$, the assignment of a single BZ is not possible due to a band crossing in the adiabatic mapping process between the 6th and 7th band. Instead the atoms share the 6th and 7th BZ with emphasis on the 6th BZ. This is illustrated by Fig.1(e), where the shapes of the 6th (blue and purple) and 7th (orange) BZs are sketched. One recognizes in Fig.1(b),(c), and (d) that the BZs are not evenly populated but rather a large fraction of the atoms resides at well localized momenta, which for the case of Fig.1(c) are highlighted by the eight red circles at the inner boundary of the 6th BZ in Fig.1(e). This localization in momentum space is more easily observed in a 3D representation, which for the case of Fig.1(c) is shown in Fig.1(f). As is substantiated below, a collision driven condensation process sets in on a timescale comparable to that of the band population procedure, which acts to redistribute the atoms into the energy minima of their band. Collision aided band decay, repopulating the 1st BZ, is only observed on a significantly longer timescale. 

In the remainder of this article we focus on the situation in Fig.1(c), which corresponds to the 7th band (the lowest of the four possible $F$-bands). In Fig.1(g) we show a plot of the 7th band derived from a band calculation involving a Fourier expansion of the Bloch-functions with 11 harmonics in each dimension and the potential of Eq.(1). The parameters $\theta/\pi = 0.66$ and $V_0/E_{\textrm{rec}} = 8.3$ correspond to the final settings in Fig.1(c). Indicated in Fig.1(g) by black disks, four local minima of the energy surface arise at the quasi-momenta $\textbf{K}_{(1,1)}, \textbf{K}_{(-1,-1)}, \textbf{K}_{(1,-1)}, \textbf{K}_{(-1,1)}$, where $\textbf{K}_{(\nu,\mu)} \equiv \frac{1}{2} \hbar k \, (\nu\, \hat x + \mu\, \hat y)$ with integers $\nu,\mu$ and $\hat x,\hat y$ denoting the unit vectors in $x$- and $y$-directions. Condensation is expected to occur at these points, which are found to coincide in their energies to better than $10^{-3} E_{\textrm{rec}}$ for a wide range of settings of $\epsilon$ and $\eta$ in Eq.(1). In particular, a local imbalance of the available standing wave intensities, which is unavoidable due to the use of finite sized Gaussian beams, does not notably lift the degeneracy. Our band calculation shows that, when  $V_0$ is ramped to zero, a crossing between the 7th and the 6th band occurs for quasi-momenta in the vicinity of the condensation points $\textbf{K}_{(\pm 1,\pm 1)}$. Thus, our experimental band mapping procedure used to obtain the picture in Fig.1(c) maps these points into the 6th rather than into the 7th BZ. This is shown in the center of the BZ illustration in Fig.1(e) where the 6th BZ is mapped onto the first BZ via translations with reciprocal lattice vectors (indicated by the red dashed arrows). Note, that the points of increased population in the measured BZ-plot in Fig.1(c) (corresponding to the red circles in Fig.1(e)) are mapped into the condensation points of Fig.1(g) (black disks).

\begin{figure}
\includegraphics[scale=0.42, angle=0, origin=c]{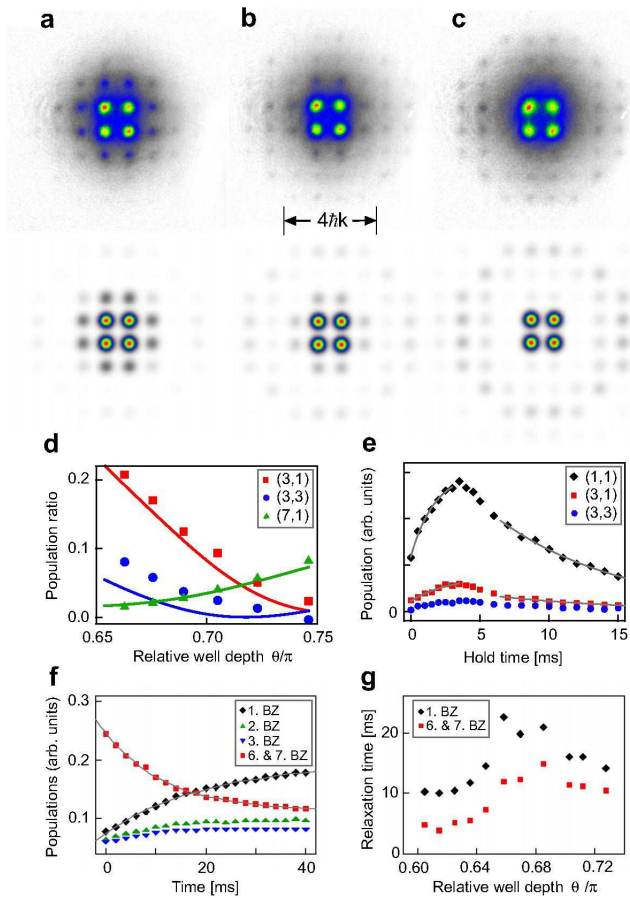}
\caption{\label{Fig.2} In (a), (b), and (c) momentum spectra are shown for a hold time of 1 ms, $V_0/E_{\textrm{rec}} = 8.3$ and $\theta/\pi = 0.66, 0.70, 0.75$, respectively. Observations (calculations) are shown in the upper (lower) row. (d) Population ratio for higher order and zero order Bragg peaks plotted versus $\theta$. Observations (calculations) are shown by the symbols (solid lines). (e) Temporal evolution of the populations of Bragg peaks. The solid lines are exponential fits applied in the wings of the graphs to determine relaxations times. (f) Temporal evolution of band populations for $V_0/E_{\textrm{rec}} = 8.3$ and $\theta/\pi = 0.67$. The solid lines are exponential fits. (g) $1/e$-times for depopulation of the 6th and 7th BZ and repopulation of the 1st BZ plotted versus $\theta$.}
\end{figure}

In Fig.2(a), (b) and (c) momentum spectra are shown for a hold time in the lattice of 1 ms, a well depth parameter $V_0/E_{\textrm{rec}} = 8.3$ and $\theta/\pi = 0.66, 0.70, 0.75$, respectively. The upper row shows the experimental observations obtained by rapidly ($< 1 \mu$s) disabling the lattice potential and absorption imaging after 30 ms of ballistic expansion. The presence of sharp Bragg peaks at quasi momenta $\textbf{K}_{(\nu,\mu)}$ for odd integers $\nu,\mu$ clearly demonstrates cross-dimensional coherence. One recognizes the absence of a zero momentum component ($\nu=\mu=0$). In fact, the lowest order components arise at the four quasi-momenta $\textbf{K}_{(\pm 1,\pm 1)}$ that we already identified as the condensation points in Fig.1(c). In the lower row the observations are contrasted with calculations of the Fourier components of the coherent superposition $\Psi_{\textrm{F}} \equiv \phi_{\textbf{K}_{(1,1)}} + i\, \phi_{\textbf{K}_{(1,-1)}}$ of the real-valued Bloch functions corresponding to the two inequivalent condensation points. The specific choice of the relative phase "$i$" will be justified below. A remarkable agreement with the observations arises with regard to the $\theta$ dependence, which is studied more quantitatively in Fig.2(d). In this graph the symbols show the population ratios between higher order ($(\nu,\mu) = (3,1),(3,3),(7,1)$) Bragg peaks and the zero order Bragg peak ($(\nu,\mu) = (1,1)$) observed for varying values of $\theta$. The peak populations are obtained by counting the number of atoms in a small circular region covering an individual Bragg peak and subtracting the number of atoms found in a surrounding ring-shaped area of the same size. The theoretical curves (solid lines) do not involve any free parameters. In Fig.2(e) we study the timescales for the formation and decay of coherence by evaluating momentum spectra at fixed values $V_0/E_{\textrm{rec}} = 8.3$ and $\theta/\pi = 0.66$ for different holding times. The graph shows the Bragg peak populations (obtained as in (d)) for the peaks identified in the inset. During the first few ms, which reflects the collisional timescale for the condensation process and the tunneling time, all peak populations increase before decay sets in with a timescale corresponding to the collisional relaxation of the band populations. Both timescales are determined by exponential fits as 1.8 ms and 9.1 ms for the zero order peak (black diamonds) and 2.9 ms and 10.1 ms for the first order peak identified by red rectangles. 

We also investigated the temporal evolution of the band populations. To this end the 7th band was populated as explained above and after a variable hold time band mapping was applied in order to produce images as in Fig.1(c). The populations found in the different BZs are plotted versus the hold time in Fig.2(f) for $V_0/E_{\textrm{rec}} = 8.3$ and $\theta/\pi = 0.67$. The 4th and the 5th BZ, which maintain nearly constant population during the observation time, have been omitted for better legibility. The graph shows a complex population redistribution, attributed to collisional dynamics. As a consequence, the initially prepared population in the 6th and 7th BZ decays and the 1st BZ is repopulated. The corresponding relaxation times are determined by means of exponential fits (solid lines) to be 12.3 ms and 19.9 ms, respectively. Similarly, the relaxation times for other values of $\theta$ are determined and plotted in Fig.2(g). The difference of the relaxation times reflects the fact that the 1st BZ is mostly populated via transfer from lower lying bands with life times longer than that of the 7th band. One recognizes, that around $\theta/ \pi \approx 0.68$ the relaxation times attain maximal values. The local $S$-orbits of the wave function $\Psi_{\textrm{F}}$ (a detailed discussion of its geometry follows below)) become maximally developed for this setting of $\theta$. Because the ground state wave function has vanishing amplitude in the shallow wells, its overlap with $\Psi_{\textrm{F}}$ is reduced for this case, and thus collisional decay into the ground state is inhibited.

\begin{figure}
\includegraphics[scale=0.3, angle=0, origin=c]{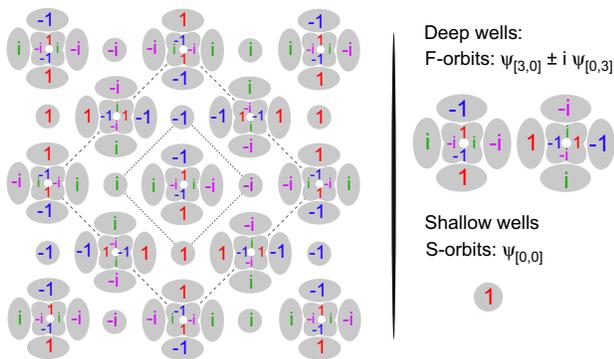}
\caption{\label{Fig.3} Orbit configuration of order parameter $\Psi_{\textrm{F}}$. The grey areas characterize the antinodal structure of the orbits, the colored numbers indicate the local phases.}
\end{figure}

The wave function $\Psi_{\textrm{F}}$ employed in the calculations of the momentum spectra has remarkable properties illustrated in Fig.3, where its local amplitude (grey areas indicate large amplitude) and phase (colored numbers) is sketched. In the deep wells $\Psi_{\textrm{F}}$ closely resembles the superposition $\psi_{[3,0]} \pm i \, \psi_{[0,3]}$ of eigenfunctions $\psi_{[n,m]}$ of a 2D harmonic oscillator with $n$, $m$ oscillator quanta in $x$- and $y$-directions. The spatial shape of $\Psi_{\textrm{F}}$ around the center of the deep wells is thus approximated by $(2x^3-3x) + i\,(2y^3-3y)$. In contrast, in the shallow wells, $\Psi_{\textrm{F}}$ mimics the harmonic oscillator $S$-orbit $\psi_{[0,0]}$. The checkerboard-like arrangement of $S$-orbits and $F$-orbits with alternating angular momentum provides equal local phases on both sides of the tunneling junctions and thus maximizes the tunneling efficiency. The inner and outer dashed rectangles denote the unit cells of the lattice and of $\Psi_{\textrm{F}}$, respectively. The sublattice of shallow wells possesses a pattern of staggered vortical currents commensurate with its plaquette structure, which match with the alternating orbital currents in the deep wells. As a consequence $\Psi_{\textrm{F}}$ breaks the translation symmetry of the lattice and time-reversal symmetry. Order parameters with similar properties have been recently predicted in the ground states of driven optical lattices \cite{Hem:07, Lim:08}.

The observed condensation momenta imply that the underlying state should be approximated by a coherent superposition or an incoherent mixture of the Bloch functions $\phi_{\textbf{K}_{(1,1)}}$ and $\phi_{\textbf{K}_{(1,-1)}}$. If coherence is assumed, and thus a wave function $\phi_{\textbf{K}_{(1,1)}}+\, c\, \phi_{\textbf{K}_{(1,-1)}}$ with some complex constant $c$, immediately $c \approx i$ follows in order to reproduce the observed momentum spectra. The setting $c = i$ maximizes the local angular momentum in the deep wells and thus minimizes the mean field energy. The reason is that in this case the local $F$-orbits acquire a maximally isotropic shape, such that the atoms (which interact repulsively) can best avoid each other. We thus conclude that, in presence of repulsive collisions, $\Psi_{\textrm{F}}$ represents the true ground state of the $F$-band. For $P$-orbits a similar prediction is discussed in Refs. \cite{Isa:05, Liu:06}. The assumption of an incoherent mixture of phases $\phi_{\textbf{K}_{(1,1)}}$ and $\phi_{\textbf{K}_{(1,-1)}}$ would necessarily imply spatial separation, since at the shallow wells both Bloch functions share equivalent local $S$-orbits. A phase separation scenario, however, requires excess kinetic energy at the phase boundaries due to inhibited tunneling and excess mean field energy. A compensating local anisotropy, sufficient to enforce local condensation in a single condensation point, is not available in our experiment. Phase separation thus appears energetically unfavorable. 

\begin{acknowledgments}
We are grateful to C. Morais Smith and L.-K. Lim for fruitful discussions. This work was partially supported by DFG (He2334/10-1, GrK 1355) and Joachim Herz Stiftung.
\end{acknowledgments}

\end{document}